\documentclass[prl,twocolumn,showpacs,amsmath,amssymb,floatfix,superscriptaddress]{revtex4}

\usepackage{graphicx}
\usepackage{color}

\usepackage[colorlinks,bookmarks=false,citecolor=blue,linkcolor=blue,urlcolor=blue]{hyperref}
\begin{document}
\title{Spontaneous trimerization in a bilinear-biquadratic $S=1$ zig-zag chain}
\author{Philippe Corboz}
\affiliation{Institut f{\"u}r Theoretische Physik, ETH Z{\"u}rich, CH-8093 Z{\"u}rich, Switzerland}
\author{Andreas M. L\"auchli}
\affiliation{Institut Romand de Recherche Num{\'e}rique en Physique des Mat{\'e}riaux (IRRMA), CH-1015 Lausanne, Switzerland}
\author{Keisuke Totsuka}
\affiliation{Yukawa Institute for Theoretical Physics, Kyoto University, Kitashirakawa Oiwake-Cho, Kyoto 606-8502, Japan}
\author{Hirokazu Tsunetsugu}
\affiliation{Institute for Solid State Physics, University of Tokyo, Kashiwa, Chiba 277-8581, Japan}
\date{\today}
\pacs{75.10.Jm, 
 75.10.Pq, 
75.40.Cx, 
75.40.Mg 
}

\begin{abstract}
Recent theoretical studies raised the possibility of a
realization of spin nematic states in the $S=1$ triangular
lattice compound NiGa$_2$S$_4$. We study the bilinear-biquadratic
spin 1  chain in a zig-zag geometry by means of the
density matrix renormalization group method and exact
diagonalization. We present the phase diagram focusing
on antiferromagnetic interactions. Adjacent to the known
Haldane-double Haldane and the extended critical phase with
dominant spin nematic correlations we find a \emph{trimerized} phase
with a nonvanishing energy gap. We discuss results for
different order parameters, energy gaps, correlation functions,
and the central charge, and make connection to field theoretical
predictions for the phase diagram.
\end{abstract}
\maketitle

{\em Introduction ---}
Quantum spin systems have provided a very wide
playground for the quest of novel quantum orders,
and the short catalog includes Haldane gap, dimer order,
chiral order and others.  Recently the discovery of spin liquid
like behavior in the spin-1 triangular magnet NiGa$_2$S$_4$ \cite{Nakatsuji05}
has stimulated increasing interests in another type of
quantum order: spin nematic order.  This is the long
range order of quadrupole moments of local spins,
in contrast to dipole moment order parameter in
conventional magnetic long range orders.
In a spin nematic ordered phase, although spins show no
static moment, spin rotation symmetry is spontaneously
broken due to anisotropic spin fluctuations.
For $S$=1 spin operators, anisotropic spin fluctuations
still have uniaxial symmetry, and this symmetry axis
is called director, in analogy to a liquid crystal.
Ferro and antiferro spin nematic
orders were independently proposed by different groups
as an explanation for the unusual low-temperature properties
of NiGa$_2$S$_4$ \cite{Tsunetsugu06, Laeuchli06b, Bhattacharjee06}.  
In particular, the antiferro spin
nematic order is possible to match the triangular lattice
structure without any frustration, and the ground state
is unique aside from degeneracy due to global spin rotation.
Although it remains open if antiferro spin nematic
order is realized in this material, it is
interesting and also important to investigate further
this state and obtain a better understanding.
To this end, we shall study a one-dimensional analog
with the same three sublattice structure, a zig-zag
chain.

\begin{figure}[t]
\center
\includegraphics[width=0.40 \textwidth]{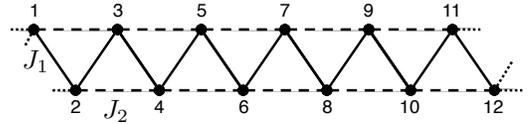} 
\caption{Zig-zag geometry of the model with $J_1$ ($J_2$) the nearest (next nearest) neighbor coupling strength.}
\label{geometry.fig}
\end{figure}

{\em Model ---}
The Hamiltonian of the bilinear-biquadratic (BLBQ) spin 1  zig-zag chain is given by
\begin{eqnarray}
\label{eqn:Hamiltonian}
H &=& J_1\  \sum_i \cos\theta ( \bf{S_i} \cdot \bf{S_{i+1}}) + \sin\theta(\bf{S_i} \cdot \bf{S_{i+1}}) ^2  \nonumber\\
 &+&   J_2\  \sum_i \cos\theta (\bf{S_i} \cdot \bf{S_{i+2}}) + \sin\theta(\bf{S_i} \cdot \bf{S_{i+2}}) ^2 .
\end{eqnarray}
$\bf{S_i}$'s are spin 1 operators and $\theta$ parametrizes the strength of bilinear and biquadratic coupling. For $\theta=0$ ($\theta=\pi/2$) the biquadratic terms (bilinear terms) vanish. For $\theta=\pi/4$ the Hamiltonian exhibits SU(3) spin symmetry. $J_1$ and $J_2$ are the nearest and next nearest neighbor coupling strengths, respectively. The model is best visualized in a zig-zag geometry where the $J_1$ bonds couple two chains and the $J_2$ bonds are located along the chains (Fig.~\ref{geometry.fig}). We concentrate on antiferromagnetic interactions on all bonds with $0 \le \theta \le \pi/2$ and $J_1,J_2 \ge 0$.
By setting $\theta=0$ the model reduces to the $J_1 - J_2$ spin 1  chain which exhibits a first order transition between two distinct topological orders \cite{Kolezhuk02}. The Haldane phase for $J_2/J_1<\alpha_t\approx0.775$, which also contains the well studied spin 1  chain for $J_2=0$, is characterized by a finite string order parameter (SOP). The ground state is a valence bond solid (VBS) where each site  consists of two spin 1/2 forming a triplet and two spin 1/2 on neighboring sites couple to a singlet. The transition into the double Haldane (DH) phase corresponds to a decoupling of the VBS string into two intertwined substrings, where each string exhibits string order \cite{Kolezhuk02}, leading to a finite \textit{double} string order parameter (DSOP).

The line $J_2/J_1=0$ corresponds to the BLBQ  chain studied, e.g., in \cite{Laeuchli06}. The gapped Haldane phase (including the Affleck-Kennedy-Lieb-Tasaki (AKLT) point for $\tan \theta=1/3$) ends at the SU(3) symmetric point $\theta=\pi/4$ (exactly solvable Uimin-Lai-Sutherland model) where a second order phase transition of Berezinskii-Kosterlitz-Thouless (BKT) type occurs~\cite{Itoi97}. Thus there is an exponentially slow opening of the gap on the Haldane side. For $\pi/4 \le \theta<\pi/2$ one finds an extended critical phase with soft modes at $k=0, \pm 2\pi/3$ and central charge $c=2$.  The dominant correlations in this phase away from the SU(3) point are quadrupolar (spin nematic)~\cite{Itoi97,Laeuchli06}.

{\em Numerical simulations ---}
The numerical results have been obtained by the density matrix renormalization group (DMRG)
\cite{White92, Schollwock05} 
method and exact diagonalization (ED). For the DMRG calculations we used up to $m=2000$ states with typically 6 sweeps and system sizes with open boundary conditions up to $L=300$ sites. 
Quantities of interest are extrapolated in $m$ and if needed also for $L\rightarrow \infty$. 
Error bars are estimated from the convergence behavior in $m$. 
With ED we considered systems with periodic boundary conditions up to $L=21$.

{\em The overall phase diagram ---}
\begin{figure}[t]
\center
\includegraphics[width=0.95\linewidth]{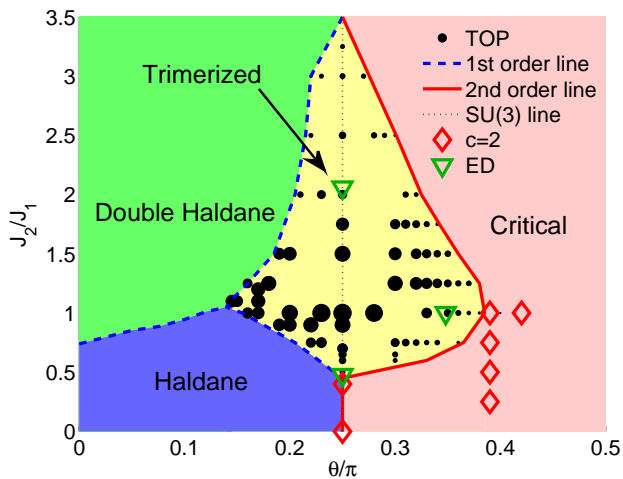}
\caption{
(Color online) The phase diagram of the spin 1 BLBQ  zig-zag chain. The area of the black dots in the trimerized phase scales with the magnitude of the trimer order parameter. We verified that the central charge is 2 in the critical phase for several points (red diamonds). The green triangles mark the phase boundary of the trimerized phase obtained from a level spectroscopy analysis from the ED data.} 
\label{phasediagram.fig}
\end{figure} 
Figure \ref{phasediagram.fig} summarizes the phase diagram of the Hamiltonian (\ref{eqn:Hamiltonian}) obtained by DMRG and ED simulations. 
We reproduced the results for the phases on the axes: Haldane, the DH, and the extended critical phase with $c=2$. 
All these phases extend into the plane. Interestingly they all touch the dominant phase in the center,
a gapped trimerized phase, which will be discussed below. 

The Haldane - double Haldane transition \cite{Kolezhuk02} point extends as a first order line in the parameter space which terminates upon touching the boundary of the trimerized phase. 
We confirmed the first order nature of the transitions by calculating the SOP and DSOP along several cuts for fixed $\theta$ and varying $J_2/J_1$.

{\em Trimerized phase ---}
The most exciting feature of the phase diagram is the gapped, trimerized phase, where three neighboring spins couple predominantly to a singlet. The trimer ground state is threefold degenerate and breaks translational invariance. This phase -- including the frustration process leading to it -- is reminiscent of the dimerized phase of the $J_1$-$J_2$ spin $1/2$  chain for $J_2/J_1\ge 0.2411$ \cite{Okamoto92}. 

Initially a massive trimerized phase for the spin 1 Heisenberg chain ($J_2/J_1=0$) for $\pi/4<\theta<\pi/2$ was put forward \cite{Nomura91, Xian93}, but later works \cite{Fath91, Reed94, Bursill95, Itoi97, Schmitt98, Laeuchli06} showed that the region remains massless and has dominant nematic correlations. In our model the additional next nearest neighbor coupling $J_2$ allows us to stabilize the trimerized state.

In previous work \cite{Penc02,Greiter07} parent Hamiltonians have been constructed using complicated four site interactions, which exhibit exact trimer ground states (in Ref.~\cite{Solyom00} yet a different kind of trimerized state was constructed). 
We now have found a trimerized phase that is stable in a finite region of the parameter space of a much simpler and possibly realistic spin Hamiltonian. We expect our model (\ref{eqn:Hamiltonian}) to be related to these parent Hamiltonians in Refs. \cite{Penc02} and \cite{Greiter07} in the same spirit as the Heisenberg spin 1 chain is related to the AKLT model.

\begin{figure}[b]
\center
\includegraphics[width=0.9\linewidth]{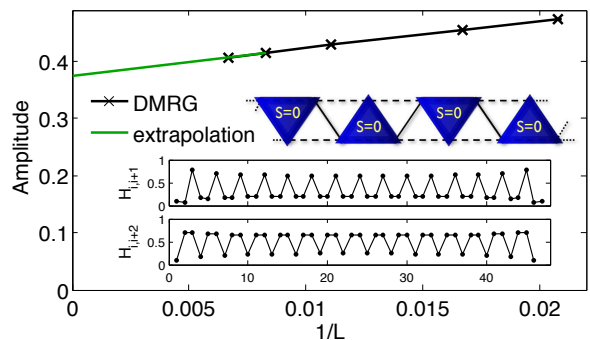} 
\caption{
(Color online) Extrapolation of the bond energy oscillation amplitude in the middle of the chain  leading to a finite trimer order parameter in the trimerized phase ($\theta=0.28\pi$, $J_2/J_1$=1).
Inset: The local bond energies form a pattern of period 3 ($L=48$ in this example). 
}
\label{bondconv_inset.fig}
\end{figure}

One of the most direct indications for a trimerized phase is the period 3 in the local bond energies (insets of Fig.~\ref{bondconv_inset.fig}). This pattern is formed because the two bonds belonging to a trimer have a lower energy than the bonds connecting two trimers.  We determine this oscillation amplitude of the local bond energies in the middle of the chain for different system sizes $L$ and extrapolate it to $L\rightarrow \infty$. In the Haldane (respectively DH) phase the amplitude vanishes exponentially with $L$. In the critical region the amplitude extrapolates to zero with a power law. But in the trimerized phase the extrapolation of the oscillation amplitude clearly yields a finite value, which we call the trimer order parameter (TOP) (see Fig.~\ref{bondconv_inset.fig}).

The magnitude of the TOP in the trimerized phase is proportional to the area of the black dots in Fig.  \ref{phasediagram.fig}. On the SU(3) line DMRG predicts that the trimerized phase sets in at $J_2/J_1\approx 0.45$ and ends at $\approx 3.5$. We used a level spectroscopy analysis of the ED data to complement the results (green triangles in Fig.~\ref{phasediagram.fig}). This technique has been successfully applied in the case of the spin 1/2 chain to locate the transition point into the dimerized phase \cite{Okamoto92}. 
ED yields a consistent result with DMRG for the lower boundary of the trimerized phase, but it suggests a substantially smaller value for the upper boundary. This mismatch stems partly from strong finite size effects on the location of the level crossings (which are much stronger than in the spin 1/2 case). 
We observe that the ED phase boundary shifts to slightly larger $J_2$ values upon taking bigger system sizes into account, approaching somewhat 
the DMRG results.
A different source of error is the rather slow convergence of DMRG in $m$ in or close to a critical phase, especially for large $J_2/J_1$.
The important result, however, which we get from both methods is the {\em finite} extent of the trimerized phase on the $J_2/J_1$ axis, in contrast to the spin 1/2 case where the dimerized phase extends to infinity. We comment on a possible explanation in the field theory section below.

\begin{figure}[b]
\center
\includegraphics[width=0.88 \linewidth]{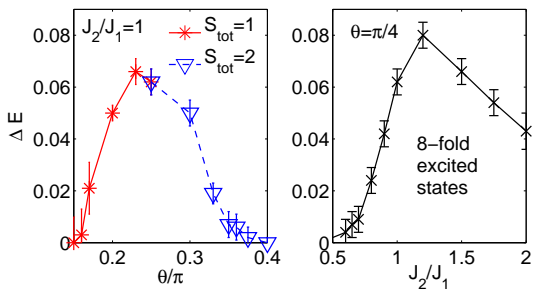} 
\caption{(Color online) Energy gaps of spin excitations along cuts in the parameter space with $J_2/J_1=1$ (left-hand side) and $\theta=\pi/4$ (right-hand side). }
\label{gaps.fig}
\end{figure}

Figure \ref{gaps.fig} shows the finite spin gap in the trimerized phase, $\Delta E = E_0(1) - E_0(0)$ where $E_0(S^z)$ is the ground state energy in the $S^z$ sector. On the SU(3) line the lowest excitation is eightfold degenerate (i.e. spin and quadrupolar excitations are symmetry related on this line). Away from the SU(3) line the lowest excitations have $S_{tot}=1$ for $\theta<\pi/4$ and $S_{tot}=2$ for $\theta>\pi/4$. According to the analysis of
 Ref.~\onlinecite{Greiter07}, the nature of excitations in the trimerized phases are gapped, deconfined domain walls, very similar to the
deconfined spinons of the spontaneously dimerized phase of the frustrated $S=1/2$ spin chain.

\begin{figure}[t]
\center
\includegraphics[width=0.9 \linewidth]{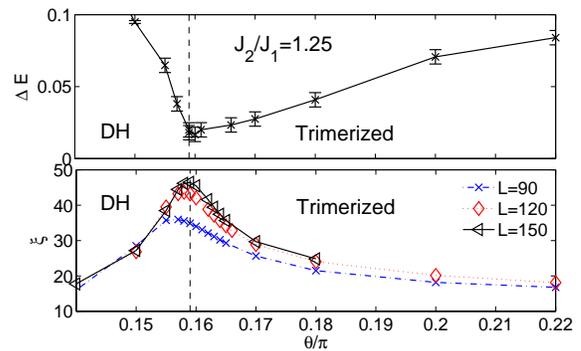} 
\caption{
(Color online) Energy gap between the ground state of singlet and triplet sector (upper plot) and spin correlation length (lower plot) for different system sizes for fixed $J_2/J_1=1.25$. The finite energy gap at the transition and the weak size dependence of the correlation length indicate a first order phase transition. The correlation length was obtained by a fit to the exponential decay of the spin-spin correlations.}
\label{dht.fig}
\end{figure}
The BKT transition line lies on the SU(3) line except for $0.45 \le J_2/J_1 \le 3.5$ where it follows the right boundary of the trimerized phase. Field theory predicts that the transition line separating the Haldane (respectively DH) phase from the trimerized phase is of first order (see below). We confirmed this numerically from a non-vanishing spin gap and a non diverging spin-spin correlation length across the transition for $J_2/J_1=1.25$ (Fig.~\ref{dht.fig}). For $J_2/J_1=0.75$ we cannot completely rule out a vanishing gap as it is rather small at the transition. Another indication for the first order nature is the jump 
of the TOP at the phase boundary, in contrast to the exponential suppression of the TOP towards the critical phase. 

\begin{figure}[t]
\center
\includegraphics[width=0.86 \linewidth]{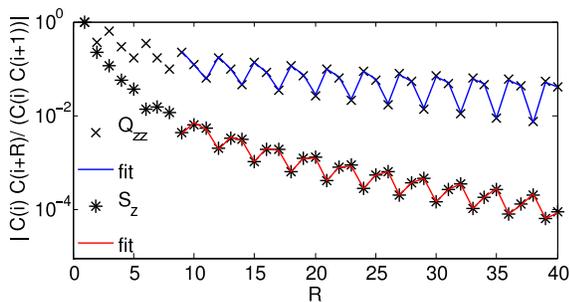} 
\caption{(Color online) Spin (stars) and quadrupolar (crosses) correlation functions (normalized) for $\theta=0.42\pi$, $J_2=0.25$. The two solid lines are fits to the functional form predicted by field theory. Logarithmic corrections to the power law lead to dominant quadrupolar correlations. }
\label{corrs.fig}
\end{figure}

{\em Critical phase ---}
We have determined the value of the central charge numerically from the
entanglement entropy for different points in the critical phase (red
diamonds in Fig.~\ref{phasediagram.fig}), and all the values agree with
$c=2$ (to within $5\%$), corresponding to a level-1 SU(3) Wess-Zumino-Witten (WZW) model \cite{Itoi97}. 
We made use of the universal scaling behavior of the entanglement entropy for conformally invariant one-dimensional quantum systems~\cite{Calabrese04} of the form 
$S(x,L) = \frac{c}{6} \ln \left[ \frac{2L}{\pi} \sin(\frac{\pi x}{L}) \right] + \mathrm{const}$,
where $S(x)$ is the von Neumann entropy of the reduced density matrix $\hat \rho(x)$ of a subsystem starting at the open boundary of length $x$ embedded in a system of length $L$, and $c$ is the central charge.
In the critical phase the spin and quadrupolar correlation functions are expected to decay as a power law with distance $x$ with an 
exponent $\eta=4/3$ and multiplicative logarithmic corrections \cite{Itoi97} $\sim \cos(2\pi x/3) (\ln x)^\sigma / x^\eta$, which suppress 
the spin ($\sigma=-2$) and enhance the quadrupolar ($\sigma=2$) correlations, such that the latter are dominant (Fig.~\ref{corrs.fig}), similar to the single chain case~\cite{Laeuchli06}.

{\em Field theoretical considerations ---}
As is discussed in \cite{Affleck-88,Itoi97}, 
the low-energy effective Hamiltonian for the model (1) 
around the point $J_{2}=0$, $\theta=\pi/4$ is the level-1 SU(3) WZW model 
perturbed by marginally irrelevant current-current 
interactions:
\begin{equation}
{\cal H}_{\text{WZW}} 
+ v\,g_{1} 
\int\!dx\!\!\!\! \sum_{A\in\text{SO(3)}} \!\!\!\!\!J^{A}_{\text{L}}J^{A}_{\text{R}} 
+ v\,g_{2} 
\int\!dx\!\!\!\! \sum_{A\in\text{others}}\!\!\!\!\! J^{A}_{\text{L}}J^{A}_{\text{R}} 
\; ,
\label{eqn:SO3-model}
\end{equation} 
where the first summation is over the SO(3)-subset of 
the SU(3)-currents $J_{\text{L,R}}^{A}$.  
The choice $g_{1}=g_{2}=g$ recovers SU(3) symmetry and the integrable 
model ($J_{2}=0$, $\theta=\pi/4$) corresponds to $g<0$.  

At the leading order, the effect of $J_{2}(>\!\! 0)$ (with $\theta=\pi/4$
fixed) may be taken into account by increasing $g(=\!\! g_{1} \!\!=\!\! g_{2})$ in Eq.
(\ref{eqn:SO3-model}).    At some critical value of $J_{2}$, 
$g$ changes its sign to positive and the system flows into 
a massive phase with three degenerate ground states \cite{Lecheminant-T-06}. 
The model (\ref{eqn:SO3-model}) has another massive phase 
characterized by the asymptotic trajectory $g_{1}=-g_{2}$,    
which has a unique (Haldane) ground state and is separated from 
the trimerized one by a first-order transition\cite{Lecheminant-T-06}.  
The Y-shaped boundaries (among {\em trimerized}, {\em Haldane} 
and {\em critical}) and other qualitative features 
in the lower portion of Fig.~\ref{phasediagram.fig} are consistent with 
what is predicted by the model (\ref{eqn:SO3-model}).  
The appearance of the gapless phase in the large-$J_{2}$ region ($J_2/J_1 \gtrsim 3.5$), as revealed by 
DMRG and ED may be understood as follows.  Let us start with two weakly coupled BLBQ chains.  
From the field-theoretical viewpoint, the ordinary two-leg ladder 
and the zig-zag chain should behave similarly in the decoupled chain 
limit; both systems will flow toward a strong-coupling limit 
upon switching on the interchain coupling $J_{1}$. 
From the standard expansion with respect to the strongly coupled rungs, 
we know that the strong-coupling phase of the ordinary two-leg BLBQ 
ladder ($\theta>\pi/4$) is critical (as in the $S=1/2$ three-leg ladder).  
Therefore we may expect the same critical behavior  
in the zig-zag ladder as well at least for sufficiently large $J_{2}/J_{1}$. 

{\em Conclusions ---}
We have presented a study of a spin 1 generalization of the famous 
$J_1-J_2$ $S$=1/2 model, motivated by recent proposals for spin nematic
ground states in a spin 1 triangular lattice~\cite{Tsunetsugu06, Laeuchli06b,Bhattacharjee06}.
The phase diagram generalizes the well-known spin fluid-dimerized transition to a 
(nematic) spin fluid to trimerized transition in the level-1 SU(3) WZW universality class. We further
explored the phase diagram in the vicinity of this transition, revealing a realization of the gapped
sector of the Andrei-Destri model. Coming from the limit of two decoupled BLBQ chains the interchain
interaction is relevant and drives a crossover to gapless single-chain behavior, in 
contrast to the $S$=1/2 case, where the marginal interaction lets the system flow to a dimerized
strong coupling state.

As a perspective we believe that the SU(N) $J_1-J_2$ model contains a general mechanism where
the critical state realized for the $J_1$ model flows towards a $N$-merized gapped state when the
ratio $J_2/J_1$ is beyond a certain critical value. Another interesting point is the possibility of stabilizing
trimerized phases in two dimensional  lattices, possibly in a $J_1-J_2$ BLBQ model on the triangular 
lattice.
Furthermore in the light of the present study, where we revealed dominant quadrupolar correlations in the
vicinity of the SU(3) line, it will be interesting to explore whether other spin nematic or 
spin-multipolar phases~\cite{Momoi06} remain to be uncovered close to SU(N) regions in higher spin antiferromagnets.
\acknowledgments
PC acknowledges inspiring discussions with M. Matsumoto, M. Troyer, and U. Schollw\"ock, and ETH Zurich 
for allocation of CPU time on the Gonzales cluster. 
AML acknowledges the support of the Swiss National Fund and thanks F. Mila and K. Penc for work on
related topics. The ED simulations have been enabled by the allocation of computing time at CSCS in Manno. HT was supported by a Grant-in-Aid for Scientific Research
(Grants No. 17071011 and No. 16540313), and also by
 the Next Generation Super Computing Project,  Nanoscience
 Program, from the MEXT of Japan.
\bibliography{spin}
\vspace{10mm}
\noindent%
\end{document}